\documentclass[notitlepage,prb, twocolumn, superscriptaddress]{revtex4-1}
\usepackage{hyperref}
\usepackage[usenames,dvipsnames]{color}
\usepackage{amsmath}
\usepackage[english]{babel}
\usepackage{amssymb}
\usepackage{bm}% bold math
\usepackage{graphicx}
\usepackage{epstopdf}
\usepackage{comment}
\usepackage{mathrsfs}
\usepackage{amsfonts}
\usepackage{bbold}
\usepackage{color,soul}
\usepackage{tabularx}
\usepackage{multirow}
\usepackage{mathtools}
\usepackage[]{todonotes}
\usepackage{algpseudocode}
\usepackage{enumitem}
\usepackage{algorithm}
\usepackage{subfig}% http://ctan.org/pkg/subfig
\usepackage{tikz}
\usepackage[compat=1.1.0]{tikz-feynman}

\begin{document}
%\linenumbers

\title{Single particle properties of the 2D Hubbard model for real frequencies at weak coupling: Breakdown of the Dyson series for partial self-energy expansions}
\author{B. D. E. McNiven}
\affiliation{Department of Physics and Physical Oceanography, Memorial University of Newfoundland, St. John's, Newfoundland \& Labrador, Canada A1B 3X7} 
\author{G. T. Andrews}
\affiliation{Department of Physics and Physical Oceanography, Memorial University of Newfoundland, St. John's, Newfoundland \& Labrador, Canada A1B 3X7} 
\author{J. P. F. LeBlanc}
\email{jleblanc@mun.ca}
\affiliation{Department of Physics and Physical Oceanography, Memorial University of Newfoundland, St. John's, Newfoundland \& Labrador, Canada A1B 3X7} 

\date{\today}
\begin{abstract}
We generate the perturbative expansion of the single-particle Green's function and related self-energy for a half-filled single-band Hubbard model on a square lattice.  We invoke algorithmic Matsubara integration to evaluate single-particle quantities for real and Matsubara  frequencies and verify results through comparison to existing data on the Matsubara axis. 
With low order expansions at weak-coupling we observe a number of outcomes expected at higher orders: the opening of a gap, pseudogap behavior, and Fermi-surface reconstruction.
Based on low-order perturbations we consider the phase diagram that arises from truncated expansions of the self-energy and Green's function and their relation via the Dyson equation.
From Matsubara axis data we observe insulating behavior in direct expansions of the Green's function, while the same order of truncation of the self-energy produces metallic behavior.  This observation is supported by additional calculations for real frequencies.  
We attribute this difference to the order in which diagrams are implicitly summed in the Dyson series.  By separating the reducible and irreducible contributions at each order we show that the reducible diagrams implicitly summed in the Dyson equation lead to incorrect physics in the half-filled Hubbard model.  Our observations for this particular case lead us to question the utility of the Dyson equation for any problem that shows a disparity between reducible and irreducible contributions to the expansion of the Green's function.
\end{abstract}

\maketitle

\section{Introduction}

Many-body perturbation theory in principle provides a window to understanding the qualitative and quantitative behavior of interacting quantum systems.  At the heart of the many-body formalism is the realization that one does not need the entire wavefunction of a system but can instead project the state of the system into contributions to the single-particle, two-particle, and ultimately many-particle observables as an expansion in the interaction order.\cite{gull:revmod,jia:2020}   
The fundamental object for single-particle properties is the single-particle Green's function $G({\bf k},\omega)$ from which one can obtain the spectral function, local density of states, and total particle density, all of which are experimental observables for condensed matter systems or cold-atom quantum emulators.\cite{kohl:eos}

The perturbative expansion is built upon the presumption that a starting non-interacting solution, $G_0$, is known.  The expansion of $G({\bf k},\omega)$ is then a functional of non-interacting Green's functions and interaction terms that may be described by a Feynman diagram.  
Evaluating diagrams to high order is computationally expensive and one recognizes that there exist repeated or factorizable structures within the full expansion of $G$. The Green's function can then be written in terms of an infinite sequence of factorizable one-particle irreducible diagrams known as the self-energy, $\Sigma$.  This gives rise to the Dyson equation which relates the non-interacting Green's function, the fully-dressed Green's function, and the self-energy function that is intended to encode all of the details of interactions at the single-particle level. 

Still, it remains difficult to sum the self-energy diagrams themselves to high order which has lead to the proposal of a variety of self-consistent or `bold-line' expansions intended to approximate infinite sums while summing only low order diagrams by repeatedly replacing the non-interacting $G_0$ with a partially-dressed  starting point for the expansion. 
While in principle this seems like a good idea, it has been shown recently that for a subset of problems based on resummation of the skeleton series, the self consistent procedure, although convergent, converges to an incorrect physical result.\cite{kozik2015nonexistence} This has been attributed to conditional convergence of the series, which allows for convergence to an arbitrary value depending upon the order in which terms are summed. 
In principle, this is not an issue for a converged expansion in that one will always obtain the correct physical result if the initial series itself has converged. This however, renders the self-consistency obsolete.

Something altogether different occurs in practice for the majority of many-body calculations.  Typically one attempts to extract physically relevant information from partial diagrammatic summations based on only very low order diagrams. In the case of diagrammatic extensions of Dynamical Mean Field Theory (DMFT) this has been shown to be dangerous in the sense that selecting a particular set of higher order contributions that can be resummed exactly produces `worse' results than a truncated low order expansion.\cite{Gukelberger:2015}  
Nevertheless, the majority of practical applications of many-body theory are based on these low-order partial diagrammatic expansions.

One example of the wide applicability of such partial expansions is the polarization function, where it is standard to use the non-interacting polarization bubble from a Lindhardt expression and then resum an  infinite chain of such diagrams, the Random Phase Approximation (RPA), as an approximation to the full susceptibility.  This is widely employed in material physics calculations based on self-consistent LDA+GW schemes.\cite{kresse:2007,kresse:2018} 
However, it has been shown rigorously that the RPA approximation while qualitatively correct, is quantitatively wrong by often an enormous margin for model systems such as the uniform electron gas\cite{chen:2019},  as well as for the 2D Hubbard model for the spin-susceptibility\cite{Schaefer:2020} when compared to the full expansions. 

The deviation of the RPA expansion from the correct answer for the full susceptibility is not unexpected given that the non-interacting bubble on which it is built explicitly neglects correlations.
Simply, if one starts by selecting a subset of diagrams that is quantitatively wrong at a given order and then creates an infinite sequence of these possibly incorrect values there should be no expectation that the end result, though well-behaved, is physically correct or relevant.  The susceptibility is therefore one example of where an infinite resummation of diagrams based on a partial diagrammatic expansion is performed, not because it is a particularly smart thing to do but purely because it is a thing that can be done.

In this paper, we consider the repercussions of this idea on the Dyson expansion of the Green's function in terms of an infinite resummation of 1-particle irreducible self-energy diagrams.  We demonstrate one case of misleading physics for the self-energy expansion for the half-filled 2D Hubbard model on a square lattice.  This is only made possible through comparison to extremely high order expansions from state of the art connected determinant method (cDET) from Ref.\onlinecite{fedor:2020} that are exact, fully-converged results in the thermodynamic limit. 

In what follows, we compute both the self-energy and Green's function expansions for the 2D square lattice Hubbard model.
We employ the method of algorithmic Matsubara integration (AMI) which allows us to evaluate each diagram both for real and Matsubara frequencies.\cite{AMI}  Access to dynamical properties on the real-frequency axis  allows us to compute directly the local density of states, in which we find the opening of a gap and reduction of quasiparticle spectral weight as expected from Matsubara axis data. 
We further find that an expansion of the Green's function for the half-filled Hubbard model, truncated at rather low order, exhibits the expected\cite{fedor:2020} metal-insulator crossover at low temperatures and weak coupling, but that truncating the self-energy at the same order only produces insulating behavior at extremely strong coupling, the latter being an observation that is in agreement with past work.\cite{yamada:4thorder} 
Together, these results suggest a fundamental flaw pervasive in studies of strongly correlated systems, namely the presumption that application of the Dyson expansion is preferable over the truncated expansion of the Green's function - when in this very fundamental weak-coupling case it is not.

\section{Model and Methods}\label{sec:modelmethods}
\subsection{Hubbard Hamiltonian}
We study the single-band Hubbard Hamiltonian on a 2D square lattice\cite{benchmarks},
\begin{eqnarray}\label{E:Hubbard}
H = \sum_{\langle ij\rangle \sigma} t_{ij}c_{i\sigma}^\dagger c_{j\sigma} + U\sum_{i} n_{i\uparrow} n_{i\downarrow} ,
\end{eqnarray}
where $t_{ij}$ is the hopping amplitude, $c_{i\sigma}^{(\dagger)}$ ($c_{i\sigma}$) is the creation (annihilation) operator at site $i$, $\sigma \in \{\uparrow,\downarrow\}$ is the spin, $U$ is the onsite Hubbard interaction, $n_{i\sigma} = c_{i\sigma}^{\dagger}c_{i\sigma}$ is the number operator, and $\langle ij \rangle$ restricts the sum to nearest neighbors for a 2D square lattice resulting in the free particle energy 
\begin{eqnarray}
\nonumber\epsilon(\textbf k)=-2t[\cos(k_x)+\cos(k_y)]-\mu,
%\nonumber\epsilon(\textbf k)=-2t[\cos(k_x)+\cos(k_y)]-4t^\prime [\cos(k_x)\cos(k_y)]-\mu.
\end{eqnarray}
where $\mu$ is the chemical potential.  Throughout we study exclusively the half-filled system which in our convention corresponds to $\mu=0$.

\subsection{Algorithmic Approach to Evaluation of Feynman Diagrams}

Algorithmic Matsubara integration is a recently developed method\cite{AMI} that allows us to evaluate the internal Matsubara sums of arbitrary Feynman diagrams symbolically at virtually zero computational expense.  
For each diagram topology of order $m$, that we denote $\zeta_m$,\cite{vanhoucke}  the amplitude is given for $d$-spatial dimensions by a Feynman integral in the form
\begin{align}
D_{\zeta_m}=\frac{U^m}{(2\pi)^{nd}\beta^n}\sum\limits_{\{\textbf k_n\}} \sum \limits _{\{\nu_n\}} &{A_{\zeta}} \prod
\limits_{j=1}^N G_0^j(\epsilon ^j, X^j ), \label{eqn:each_diag_in_goal}
\end{align}
that must be summed over a set of internal Matsubara frequencies $\{\nu_n\}$ and internal momenta $\{\textbf k_n\}$. $A_{\zeta}$ is a topology dependent prefactor and the product of $N$ non-interacting Green's functions $G_0^j$ that are each given by
\begin{equation}
{G}^j_0=\frac{1}{X^j-\epsilon^j} ,    
\end{equation}
where $X^j$ represents the frequency, and $\epsilon^j$ the energy of each function. $X^j$ and $\epsilon^j$ are constructed via symbolic representations\cite{GIT}. 
For a given diagram with $n$ independent (internal) Matsubara frequencies we define the frequency of each $G_0^j$ as the linear combination $X^j=\sum_{\ell=1}^{n+1}i\alpha_\ell^j \nu_{\ell}$, where the allowed values for the coefficients $\alpha_\ell^j$ are zero, plus one, or minus one. Similarly the free particle energy is $\epsilon^j = \epsilon(\textbf k_j)$, where $\textbf k_j=\sum_{\ell=1}^{n+1} \alpha_{\ell}^j \textbf k_{\ell} $. In this notation $\nu_{n+1}$ and  $\textbf k_{n+1}$  are the external frequency and momenta, $\nu_{ext}$ and ${\bf k}_{ext}$ respectively.

 The Matsubara summations of a given Feynman diagram are contained in the factor
\begin{align}
I_{\zeta_m}=\frac{1}{\beta^n} \sum \limits _{\{\nu_n\}} \prod
\limits_{j=1}^N  G_0^j(\epsilon ^j, X^j ).  \label{eqn:AMI_target}
\end{align}
This allows us to write the contributions from a diagram as 
\begin{align}
D_{\zeta_m}=\frac{1}{(2\pi)^{nd}}\sum\limits_{\{\textbf k_n\}} \mathfrak {D}_{\zeta_m}, \label{eqn:goal_after_AMI}
\end{align}
where
\begin{align}
\mathfrak {D}_{\zeta_m}=A_{\zeta} U^m I_{\zeta_m}(\{i\nu_{\rm ext}\},\{\textbf k_{\rm ext}\},\beta,\mu).   \label{eqn:each_diag_after_AMI}
\end{align}

AMI\cite{AMI,GIT} provides a tool to evaluate Eq.~(\ref{eqn:AMI_target}).
We first generate all diagrams of a given order for an observable of interest and then use AMI to generate an analytic expression in the form of Eq.~\ref{eqn:each_diag_after_AMI}.  For the 2D model at order $m$, this effectively resolves one third of the internal integrals leaving only internal momenta to be summed.  
The resulting integrand, $I_{\zeta_m}$, typically contains many terms and is a function of internal momenta, but also external variables, such as temperature or chemical potential, as well as external momenta and external frequencies.  The AMI result is an analytic function of the external frequency, which in turn allows us to perform  analytic continuation $i\omega_n \to \omega +i\Gamma$ without having to resort to the ill-conditioned process of numerical analytic continuation.\cite{Levy2016, jarrell:maxent, gull:nevanlinna}

AMI has disadvantages when compared to state-of-the-art determinental methods.\cite{simkovic2017determinant,moutenet2018determinant} In our approach diagrams are sampled independently and the diagrammatic space grows dramatically with expansion order and also the number of terms in the AMI result, $I_{\zeta_m}$, grows exponentially with expansion order.  
In the case of the self-energy expansion, the second order diagram (for which the Matsubara sums can be managed by hand) produces 4 terms.  For our calculations, the fourth order contribution to $\Sigma$ involves 12 diagrams for which the AMI procedure yields a total of  932 analytic terms, while at sixth order there are 515 diagrams resulting in $\sim 1.1\times 10^6$ analytic terms. The two latter cases generate an impossibly large number of terms to manage by hand and, consequently, this work is only made possible via the AMI procedure\cite{AMI} or other recently developed algorithmic or analytic schemes\cite{jaksa:2020, jaksa:analytic}. 
We note that substantial computational savings can be achieved by factorizing and grouping diagrams based on their pole structure, which at half-filling results in groups of equivalent diagrams.\cite{GIT}
While the computational expense of sampling diagrams individually is large, it affords us the opportunity to dissect contributions from individual diagram groups which cannot be done using determinental methods in which all diagrams of a given order are summed simultaneously.

\subsection{Observables}
For this work, we generate the diagrammatic expansion for the Green's function up to 6th order, and separate out the 1-particle irreducible self-energy diagrams. 
This gives us two routes to access information for single-particle properties.  The first via the direct expansion of the Green's function, that includes all 1-particle reducible and irreducible diagrams, truncated at order $n$ which we will denote $G^{(n)}$.  Alternatively, we can evaluate only the irreducible self-energy diagrams summed to order $n$,
 $\Sigma^{(n)}$, and use the Dyson equation to represent the Green's function which we will denote $G_{\Sigma^{(n)}}$ given by 
\begin{equation}
    G_{\Sigma^{(n)}}=\left( G_0^{-1}-\Sigma^{(n)}\right)^{-1}.\label{eqn:Gsigma}
\end{equation}

Conversely, the same methodology can be applied in reverse.  By first truncating the Green's function expansion we can invert the Dyson equation to suggest what the related self-energy, which we denote $\Sigma_{G^{(n)}}$, might be via
\begin{equation}
    \Sigma_{G^{(n)}}=G_0^{-1}-\left[G^{(n)}\right]^{-1}. 
    \label{eqn:sigmaG}
\end{equation}
 We therefore have two routes to obtain the Green's function: 1) the direct evaluation of $G^{(n)}$, and 2) the evaluation of $G_{\Sigma^{(n)}}$ via Dyson's equation.  Similarly, the inverse procedure provides two representations of the self energy.  
  The Green's function expansion truncated at order $n$ includes all reducible and irreducible diagrams up to order $n$ while the self-energy expansion when coupled with Dyson's equation includes all reducible and irreducible diagrams up to order $n$ \emph{and} all reducible diagrams at higher order that are comprised of the known diagrams.
 Regardless of which approach is taken the two representations are equivalent in the $n\to \infty$ limit, but for a truncated expansion these two schemes amount to summing the same diagrams of the series selected in a different order. In practice, both are obtained from the same information via computation of $\Sigma^{(m)}$ for $m\leq n $ from which the truncated expansion of $G^{(n)}$ can be constructed by including irreducible diagrams as well as all possible combinations of chains of  1-particle reducible self-energy insertions. 

This provides an important diagnostic for truncated expansions since deviation of $G^{(n)}$ from $G_{\Sigma^{(n)}}$ immediately indicates non-convergence of the series. 
With this in mind, we present results for the spectral function obtained from the expansion of the Green's function, $A({\bf k},\omega)$=$-{\rm Im}G^{(n)}({\bf k},\omega)$, along with the equivalent result from the Dyson expansion using a truncated self-energy, $A_{\Sigma}({\bf k},\omega)=-{\rm Im}G_{\Sigma^{(n)}}({\bf k},\omega)$.  
We also present results for the density of states, $N(\omega)$, which can be obtained from the expansion of the Green's function, $N(\omega)=\sum\limits_{k}A({\bf k},\omega)$, where the final momenta summation can be performed simultaneously with the internal momenta summations from the resulting AMI integrand Eq.~(\ref{eqn:goal_after_AMI}). 

In line with previous works,\cite{fedor:2020, Schaefer:2020,behnam:2020} we examine how the difference between the imaginary parts of the self-energies at the first and second Matsubara frequencies acts as an indicator of non-Fermi-liquid behavior.  We compute $\Delta\Sigma^{(n)}({\bf k})={\rm Im}\Sigma^{(n)}(i\omega_0,{\bf k})- {\rm Im}\Sigma^{(n)}(i\omega_1,{\bf k})$ as well as $\Delta\Sigma_{G^{(n)}}({\bf k})={\rm Im}\Sigma_{G^{(n)}}(i\omega_0,{\bf k})- {\rm Im}\Sigma_{G^{(n)}}(i\omega_1,{\bf k})$ for values of $\beta$ and $U/t$.  At fixed $\beta$, we sweep over values of $U/t$ to find where $\Delta \Sigma=0$, marking a crossover temperature that we denote $T^*_\Sigma({\bf k})$ and $T^*_G({\bf k})$ respectively for a particular choice of momenta, ${\bf k}$.  However, it is important to note that the use of $\Delta\Sigma$ as a metric for non-Fermi-liquid behavior is only rigorously correct in the $T\to0$ limit. 

Throughout, we present results truncated at fourth order and where possible include data at sixth order.  We emphasize that even at fourth order, the number of terms to be evaluated is extremely large and the analytic expressions have not been available prior to the advent of AMI. 

\section{Results}\label{sec:results}

\begin{figure}
 \centering
\includegraphics[width=\linewidth]{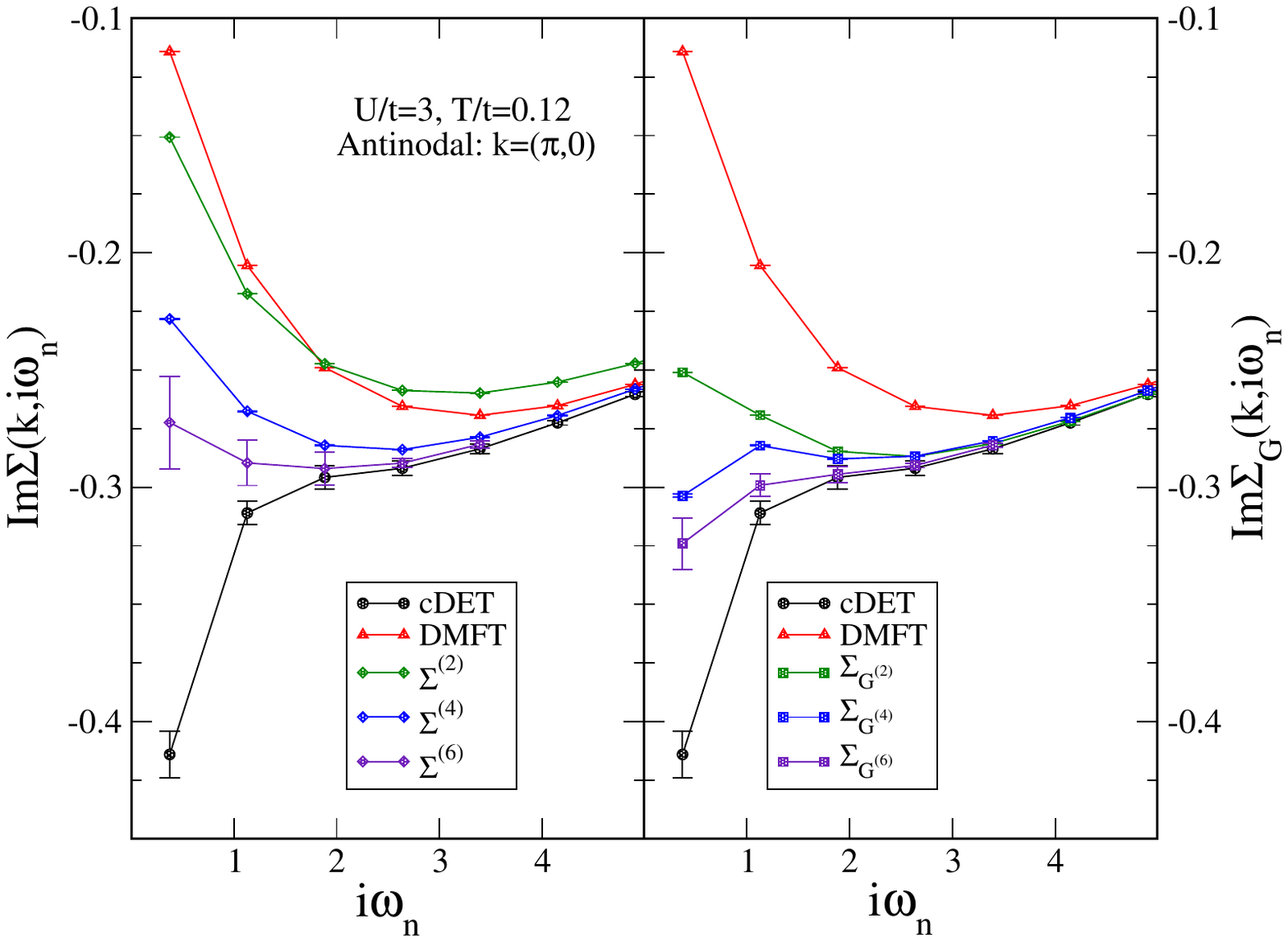}
 \caption{Second, fourth, and sixth order self-energies obtained directly ($\Sigma$, left), and from inverting Dyson's equation ($\Sigma_G$, right) at $U/t=3$, $\beta t=8.33$, $\mu=0$ for the antinodal momenta ${\bf k}=(\pi,0)$.  Shown for reference are DMFT results and benchmark results from cDET taken from Ref.~\onlinecite{fedor:2020}.}
 \label{fig:FedorSigma}
 \end{figure}

We show results for $\Sigma$ and $\Sigma_{G}$ for a well established benchmark case in Fig.~\ref{fig:FedorSigma}.  Plotted on the Matsubara axis, this data set at $U/t=3$ and $\beta t=8.33$ for the antinodal point ${\bf k}_{an}=(\pi,0)$ is known to high order from the connected determinant method (cDET)\cite{fedor:2020,rossi2017determinant}.  Paying special attention to $\Delta \Sigma={\rm Im}\Sigma(i\omega_0)-{\rm Im}\Sigma(i\omega_1)$ we see from the cDET data the characteristic negative value of $\Delta\Sigma$, suggesting that the system has non-Fermi-liquid or insulating character.
In contrast, we present DMFT results (red curves in Fig.~\ref{fig:FedorSigma}) computed using the numerically exact continuous time auxiliary field method.\cite{gull:revmod}  We see that the local approximation is insufficient at weak coupling, an issue only recently understood.\cite{fedor:2020,Schaefer:2020}

Our AMI calculation of $\Sigma^{(2)}$, $\Sigma^{(4)}$, and $\Sigma^{(6)}$ is shown in the left frame of Fig.~\ref{fig:FedorSigma}.  We see that with increasing order the result  systematically approaches the cDET benchmark. By sixth order, however, the sign change in $\Delta\Sigma$ is not evident, and each curve shows a positive $\Delta\Sigma$ suggestive of metallic behavior.  
We plot also results from the truncation of the Green's function at order n, $\Sigma_{G^{(n)}}$.  Starting at second order, $\Sigma_{G^{(2)}}$ is substantially closer to the cDET benchmark than  $\Sigma^{(2)}$, particularly at high frequencies.  At fourth order, $\Sigma_{G^{(4)}}$ lies even closer to the cDET benchmark and, more importantly, demonstrates the characteristic non-Fermi-liquid signature of $\Sigma(i\omega_0)<\Sigma(i\omega_1)$. Inclusion of sixth-order contributions to $\Sigma_{G}$ results in further improvement in the agreement between our result and the benchmark.   

This dichotomy of physical outcomes is of course irrelevant in the large $n$ limit.  We see that both $\Sigma^{(n)}$ and $\Sigma_{G^{(n)}}$ are trending towards the benchmark.  What is disconcerting is how different the two approximations are for truncation at low orders and that $\Sigma_{G^{(n)}}$  is systematically closer to the correct result than $\Sigma^{(n)}$.   This difference is in spite of the two schemes requiring  precisely the same computational effort - both obtained from knowledge of $G_0$, $\Sigma^{(2)}$, $\Sigma^{(4)}$, and $\Sigma^{(6)}$.

\begin{figure}
    \centering
    \includegraphics[width=\linewidth]{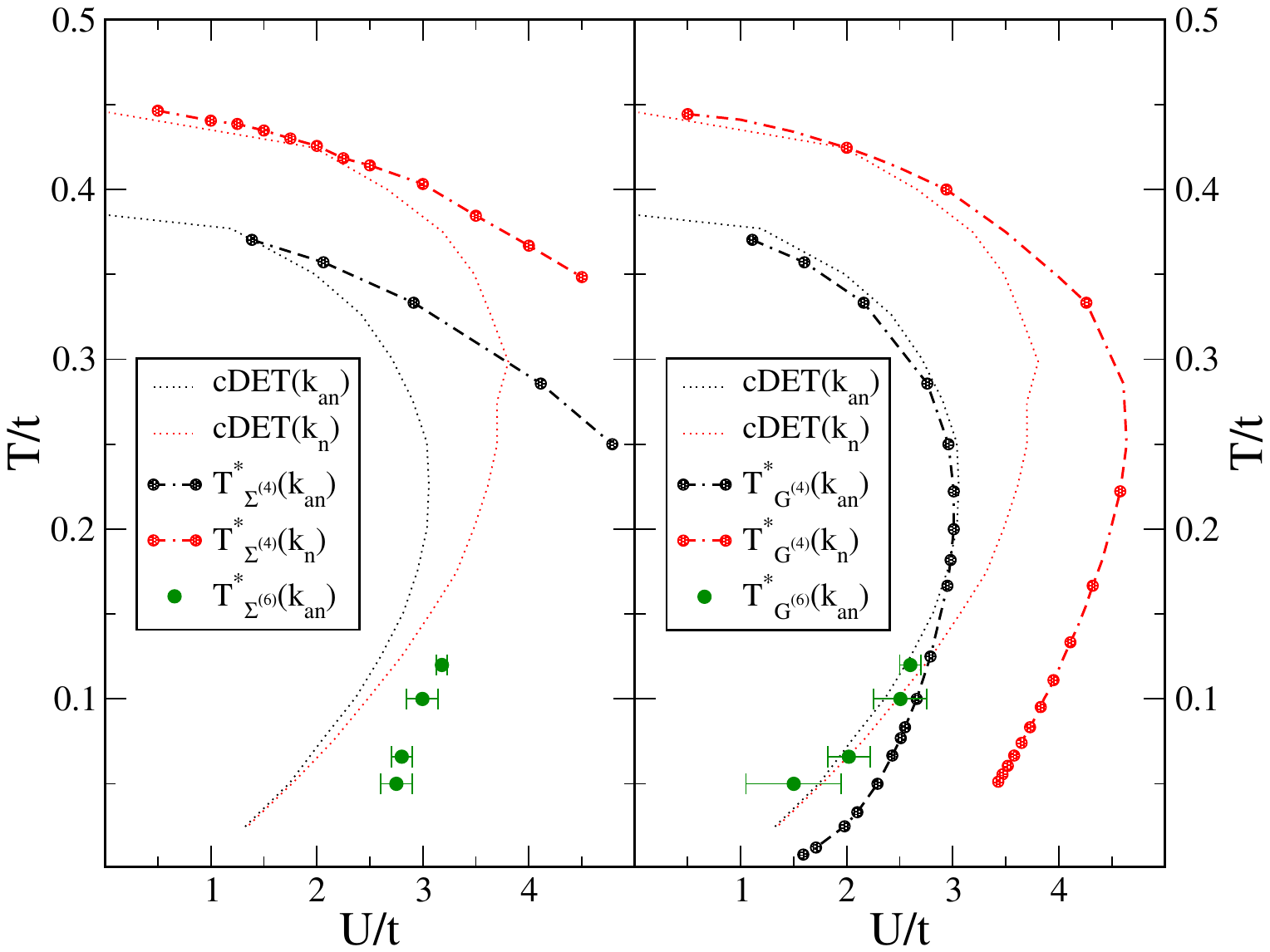}
    \caption{Phase diagram in the T-U plane for fourth, and sixth order. Left:$T^*_\Sigma({\bf k})$, and Right:$T^*_G({\bf k})$.  Benchmark data for cDET taken from Ref.~\onlinecite{fedor:2020}.  Uncertainty associated with fourth order data is the approximate size of each data point.}
    \label{fig:phase}
\end{figure}

To explore this issue further, in Fig.~\ref{fig:phase} we present  a phase diagram in the $T-U$ plane and identify the temperature at which $\Delta\Sigma$ changes sign for the antinodal and nodal momenta, ${\bf k}_{an}$ and ${\bf k}_n$, respectively. 
Also shown is cDET data from Ref.\onlinecite{fedor:2020} for nodal and antinodal momenta.  This benchmark has several key features present at both nodal and antinodal points: 1) at low temperature the nodal and antinodal crossovers merge and suggest a critical interaction strength at $T/t=0$ of $U_c< 1.3$; 2) a turning point above some temperature at a maximal value of $U/t$; 3) a high temperature intercept in the $U/t\to0$ limit.  

We truncate our calculations at fourth order and construct a phase diagram from both $\Sigma^{(n)}$ [left frame] and $\Sigma_{G^{(n)}}$ [right frame]. 
Considering first $T^*_{\Sigma^{(4)}}$,  we see that our data merges with the high-temperature crossover for small values of $U/t$.  However, unlike the cDET benchmark, $T^*_{\Sigma^{(4)}}$ does not show a turning point for either nodal or antinodal points and therefore also does not show a low temperature crossover.  This observation is consistent with the absence of a downturn in $\Sigma^{(4)}$ in Fig.~\ref{fig:FedorSigma} as well as with past studies of 4th order expansions based on Matsubara axis data.\cite{yamada:4thorder}.  
For ${\bf k}_{an}$ we also compute $T^*_{\Sigma^{(6)}}$ shown as green circles at $\beta t= 8.33$, $10$ and $20$.  We see that the sixth order results has shifted significantly toward lower values of $U/t$ - while at fourth order a crossover requires critical interaction strengths $U_c/t>6$, at sixth order $U_c/t \sim 2.6$ at our lowest temperature shown, $\beta t=20$.

The story is quite different for the same phase diagram generated from the Green's function expansion $T^*_{G^{(n)}}$.  At fourth order we see that $T^*_{G^{(4)}}({\bf k}_{an})$ precisely follows the cDET result over the range $0.37\geq T/t \geq 0.18$, below which deviations are seen between the two sets of results.  The key characteristic that is recaptured is the reentrant turning point and low temperature crossover.  This crossover shows a monotonically decreasing critical $U/t$ for decreasing temperature and persists to our lowest accessible temperature of $\beta t=80$ at $U/t\sim 1.6$. $T^*_{G^{(4)}}({\bf k}_n)$ shows behavior similar to that of $T^*_{G^{(4)}}({\bf k}_{an})$, but  with considerably larger deviation from the benchmark results.  This suggests that the convergence properties of the nodal and antinodal points may be substantially different and that the nodal point requires relatively larger-order diagrams to converge.  Finally, our calculation of $T^*_{G^{(6)}}({\bf k}_{an})$ substantially improves upon $T^*_{G^{(4)}}({\bf k}_{an})$ converging with the benchmark, here shown down to $\beta t=20$.  

From Fig.~\ref{fig:phase}, we see that the result from the direct expansion $G^{(n)}$ shows results that are systematically closer to the benchmark than results from $\Sigma^{(n)}$.  This remains true throughout the phase diagram and is consistent with the behavior seen at the single point ($U/t=3$, $\beta t=8.33$) shown previously in Fig.~\ref{fig:FedorSigma}.   Particularly, above $T/t=0.18$, the antinodal behavior at fourth order is in precise agreement with the benchmark.  
The observation of a crossover based on Matsubara axis data leads to the obvious question as to how well this phase diagram is reflected in real-frequency properties since the observation that a sign change in $\Delta \Sigma$ relates to non-Fermi-liquid behavior has  only previously been examined from Matsubara data via numerical analytic continuation.\cite{park:2008} 

\begin{figure}
 \centering
  \includegraphics[width=0.5\textwidth]{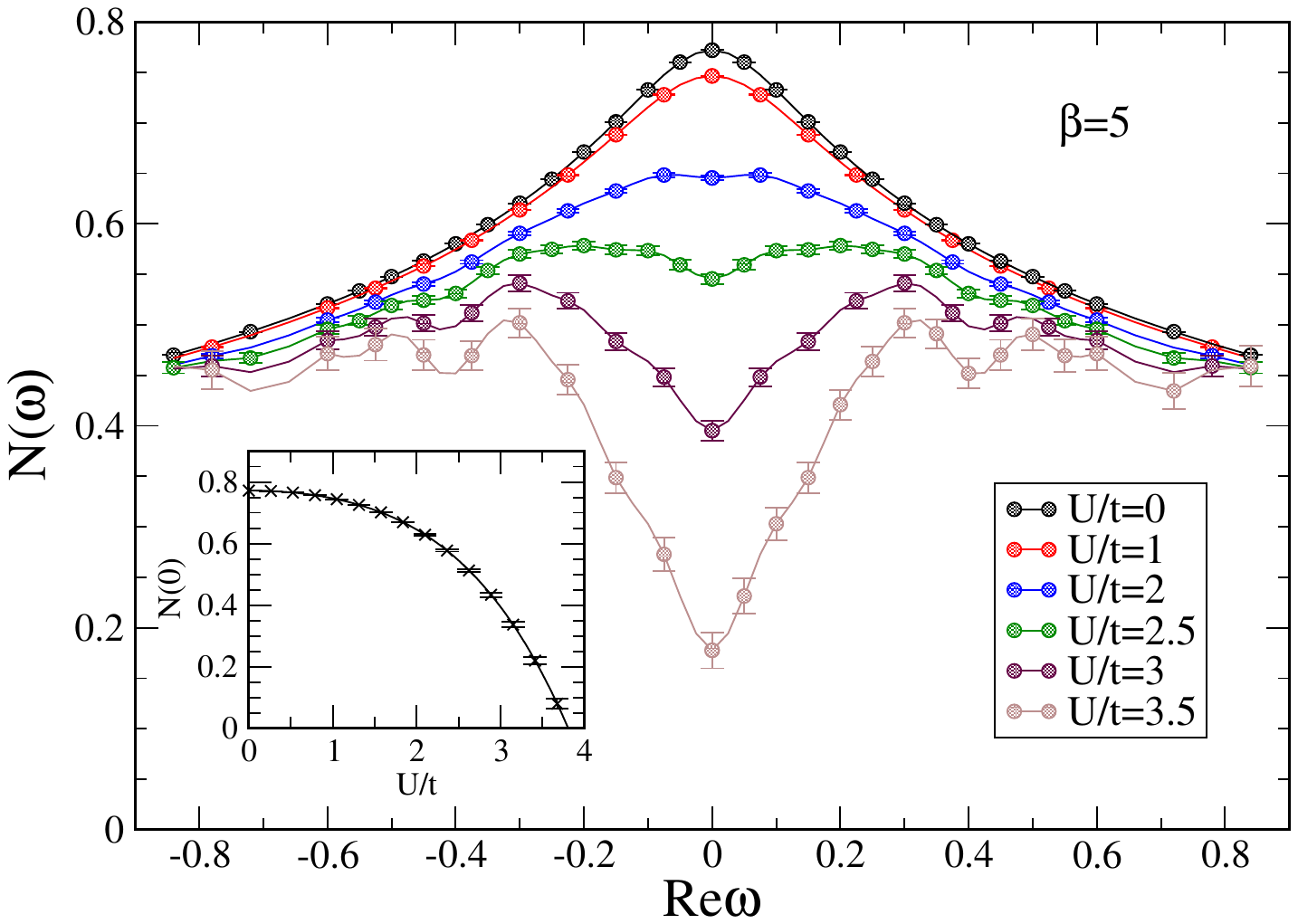}
 \caption{Density of states $N(\omega)$ for $\beta t=5$ at varying $U/t$ values. Inset shows the density of states at $\omega=0$ for $\beta t=5$.  A broadening of $\Gamma/t=0.125$ is used in the analytic continuation.}
 \label{fig:dos}
 \end{figure}

In Fig.~\ref{fig:dos}, we compute the density of states $N(\omega)$ from ${\rm Im}G^{(4)}({\bf k},\omega+i\Gamma)$ shown as a function of $\omega$ for increasing values of $U/t$.  We use a moderate broadening of $\Gamma/t=0.125$ that acts as a numerical regulator required for the integration of Eq.~(\ref{eqn:goal_after_AMI}), the overall impact of which  is to soften features smaller than $\Gamma/t$.  For example, at $U/t=0$ the value of $\Gamma$ suppresses the van Hove singularity but is otherwise inert. 
We see that as $U/t$ increases there is a suppression of states near the Fermi level at frequencies $|\omega| < 0.25$. For $U/t>3$, $N(\omega)$ exhibits a minimum at $\omega=0$ and a linear in $|\omega|$ behavior indicative of a $d$-wave gap that plateaus and exhibits a ringing or peak-dip-hump structure from $|\omega|=0.3\to 0.5$.\cite{pdh}  Such behavior is considered a sign of strong boson mediated interactions and can be related through a derivative of $N(\omega)$ to the $\alpha^2 F$ coupling amplitude\cite{carbotte:review, benneman} which in this case would be spin-fluctuations that are known to cause pseudogap behavior.\cite{gunnarsson:2015, behnam:2020}

To facilitate comparison with the phase diagram of Fig.~\ref{fig:phase}, we show in the inset of Fig.~\ref{fig:dos} the value of $N(0)$ as a function of $U/t$ at $\beta t=5$.  Comparing that data to the  Matsubara data in Fig.~\ref{fig:phase} we see that the  ${\bf k}_{an}$ crossover that occurs at $U/t=3$ corresponds well with the decrease in $N(0)$.
 From the inset we see that $N(0)$ tends to zero around $U/t=3.7$, a value only marginally higher in $U/t$ than the nodal crossover result from cDET beyond which the system should be insulating.  Overall, there is a clear correspondence between the real frequency density of states and the Matsubara axis phase diagram. 

\begin{figure}
 \centering
  \includegraphics[width=0.5\textwidth]{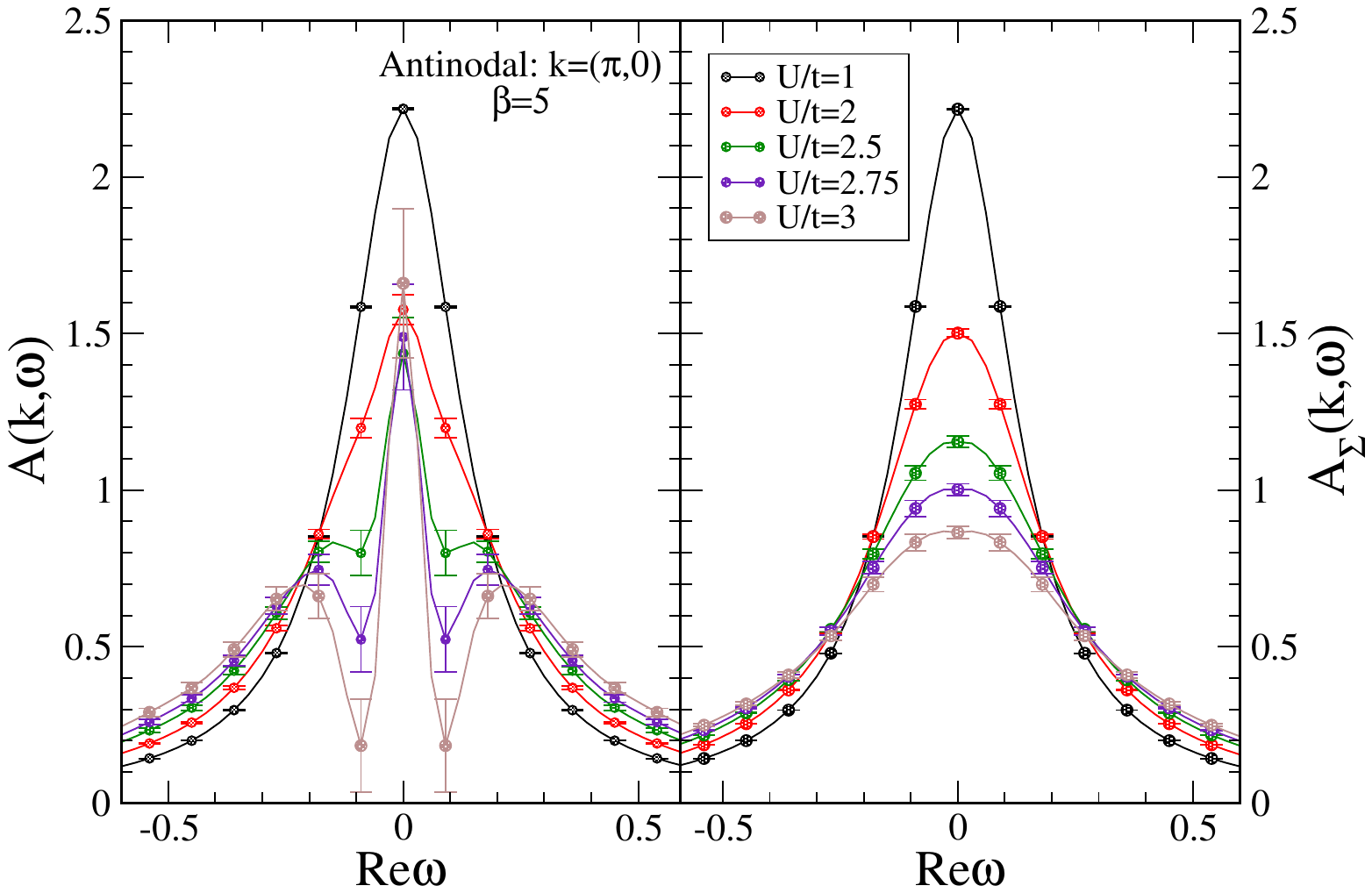}
 \caption{Spectral function at the antinodal point calculated directly from the Green's function (left, $A({\bf k},\omega)$) and from the self-energy via Dyson's equation (right, $A_{\Sigma}({\bf k},\omega)$). A broadening of $\Gamma/t=0.125$ is used in the analytic continuation. }
 \label{fig:AvsAG}
 \end{figure}

Returning to the deviation between the $G^{(n)}$ and $\Sigma^{(n)}$ expansions, in Fig.~\ref{fig:AvsAG} we now invert the comparison and plot the spectral functions from $G^{(n)}$ and  $G_{\Sigma^{(n)}}$ obtained from Eq.~\ref{eqn:sigmaG}, truncated at fourth order, on the real frequency axis at the antinodal momentum point.
$A({\bf k}_{an},\omega)$ and $A_\Sigma({\bf k}_{an},\omega)$ are plotted in the left and right frames, respectively, for increasing values of $U/t$.
At $U/t=1$, the results are identical between the two frames, suggesting that the series is converged.  
We see deviations emerging above $U/t=2$, where $A_\Sigma({\bf k},\omega)$ broadens but remains a single peak while $A({\bf k},\omega)$ begins to form a characteristic lower and upper Hubbard band with a remnant $\omega=0$ quasiparticle peak.

The data for $A({\bf k},\omega)$ and $A_\Sigma({\bf k},\omega)$ further corroborates  behaviour shown in Figs.~\ref{fig:FedorSigma} and \ref{fig:phase}, that the truncated expansion $G^{(n)}$ shows correct insulating behavior while $G_{\Sigma^{(n)}}$ does not.  This suggests that the physics contained in the two expansions is vastly different.  
Further, we emphasize that the left and right frames of Fig.~\ref{fig:dos} are not the same above $U/t=1$, suggesting that the expansion at fourth order  has not converged. While this means that neither calculation should be expected to be quantitatively correct, this does not preclude us from asking the question as to which of the two is closer to the correct result. Based on our understanding of Fig.~\ref{fig:phase}, $A({\bf k}_{an},\omega)$ from $G^{(n)}$ is consistent with the cDET reference data while $A_\Sigma({\bf k}_{an},\omega)$ is not.  
We revisit and explain this effect in Section~\ref{sec:dissect}.

\begin{figure}
    \centering
    \includegraphics[width=\linewidth]{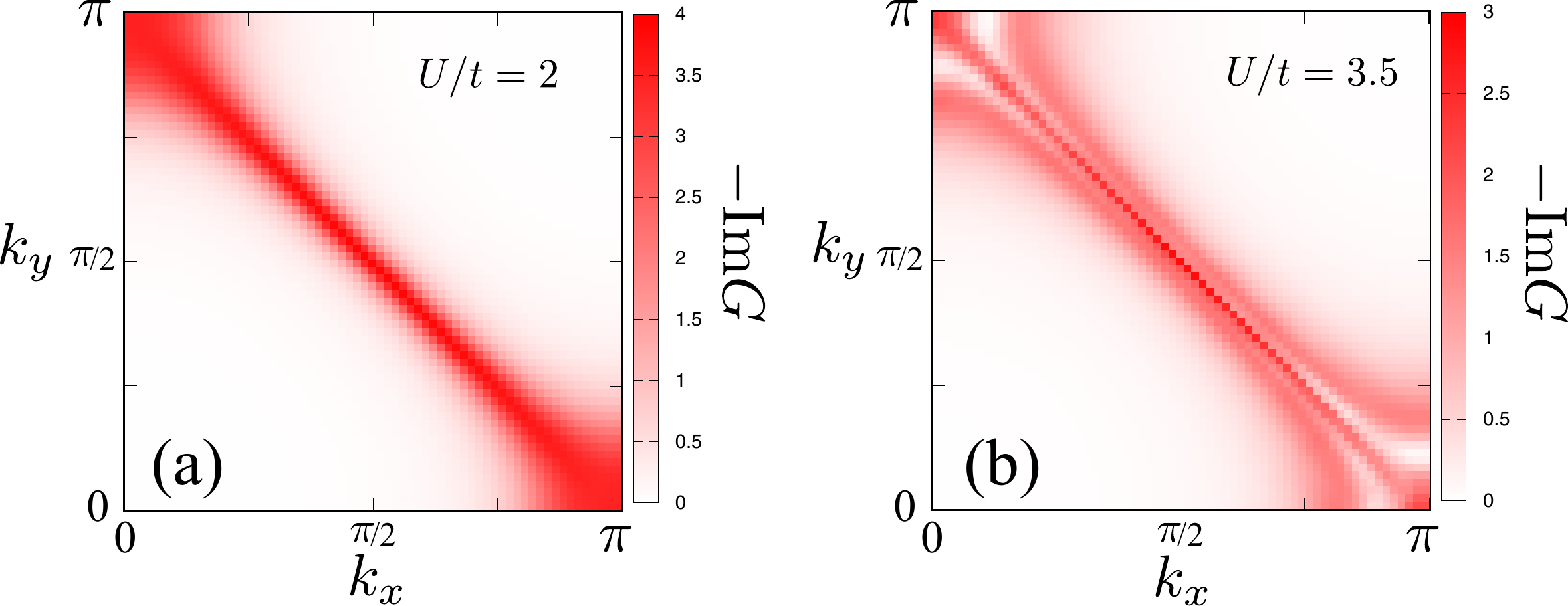}
    \includegraphics[width=\linewidth]{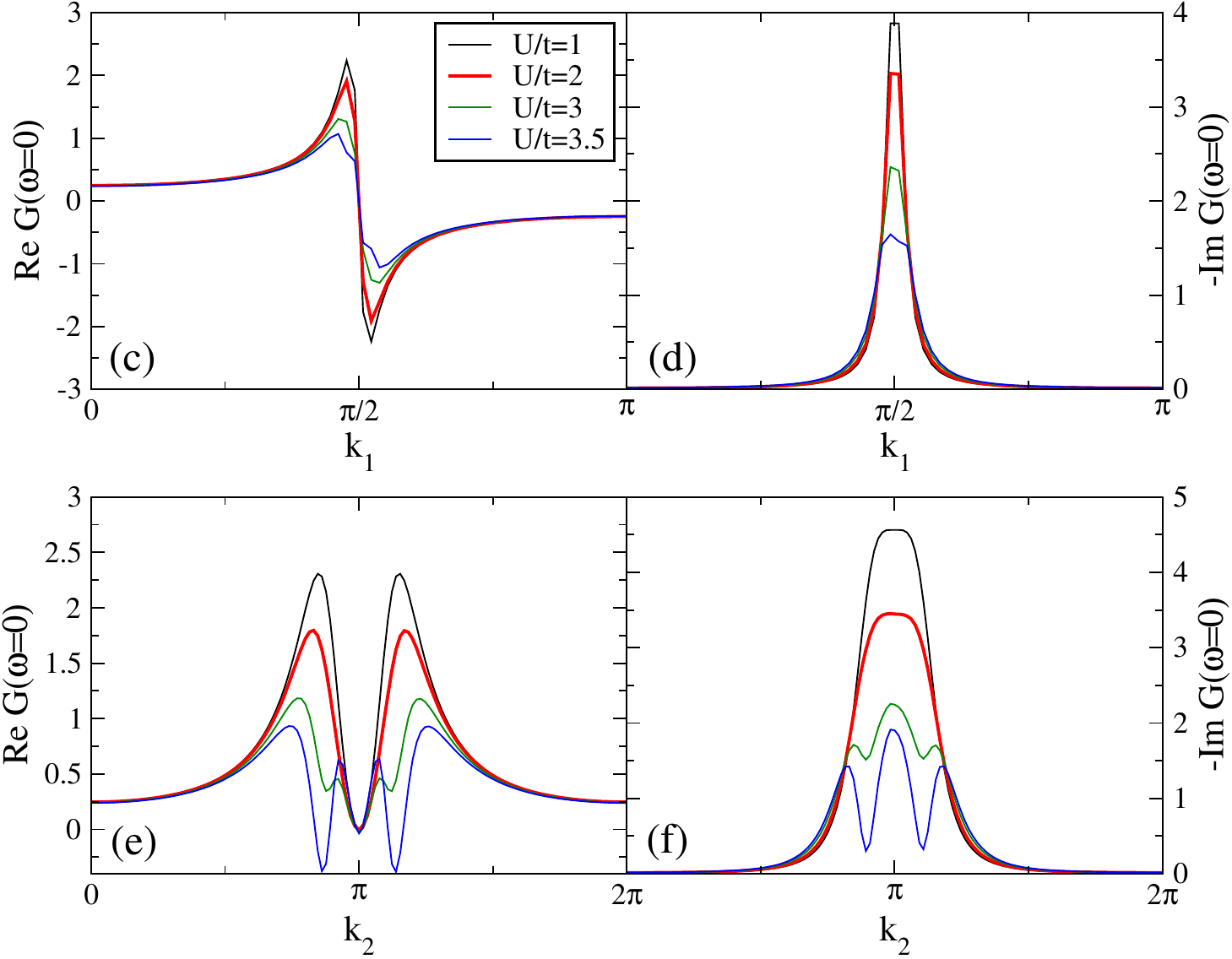}
    \caption{Top: $-{\rm Im}G(k_x,k_y,\omega=0)$ for 
    $\beta=5$ for $U/t=2$ (a) and $U/t=3.5$ (b) at half-filling $\mu=0$.
    Bottom: Real and imaginary part of the Green's function for cuts along: ${\bf k}_1: (k_x,k_y=k_x)=(0,0)\to (\pi,\pi)$ and ${\bf k}_2: (k_x=0,k_y)=(0,0)\to(0,2\pi)$. }
    \label{fig:kxkyandslices}
\end{figure}

In Fig.~\ref{fig:kxkyandslices}(a) and (b),  we consider the spectral function at the Fermi level, $\omega\to 0+i\Gamma$, as a function of ${\bf k}=(k_x,k_y)$ on a high resolution $65\times 65$ momentum grid.  Due to the large number of points we truncate the Green's function at fourth order.  We present results for two representative interaction strengths: $U/t=2$ and $U/t=3.5$.  We see that at $U/t=2$, there is a well defined Fermi surface comprised of a single spectral peak throughout the Brillouin zone. With an increase in interaction strength to $U/t=3.5$, the picture changes to include three spectral peaks, one along the original Fermi surface, and two beginning to form on either side of the original Fermi surface.  Since these peaks occur at the same energy, they should not be confused with the upper and lower Hubbard bands that form as a function of energy as in Fig.~\ref{fig:AvsAG}.  Here, the peaks represent a shift in spectral weight in momenta rather than in energy.

We also present results as a function of momentum for paths through the Brillouin zone, shown in Fig.~\ref{fig:kxkyandslices}(c)$\to$(f), for increasing values of $U/t$ at fixed temperature.  We present two cuts: the first along the nodal direction from ${\bf k}_1=(0,0)\to (\pi,\pi)$ (frames (c) and (d)), and the second along the antinodal direction from ${\bf k}_2=(0,0)\to(0,2\pi)$ (frames (e) and (f)). 
Along the ${\bf k}_1$-direction,  shown in Fig.~\ref{fig:kxkyandslices}(d), there remains a single peak in the vicinity of $(\pi/2,\pi/2)$.  We see that increased interaction strength reduces the peak amplitude but otherwise it is virtually unchanged.  The behavior of ${\rm Re}G$ in frame (c) is similarly simple showing a loss of peak amplitude and slight change of slope near $\omega=0$ and further shows a single sign change across $(\pi/2,\pi/2)$. 
In contrast, the path along ${\bf k}_2$ in Fig.~\ref{fig:kxkyandslices}(f) shows a reduction in peak amplitude as well as spectral weight followed by additional peaks at incommensurate wave-vectors above $U/t=3$.  These appear at a shift $\delta \sim 0.51$ above/below and left/right of the point $(0,\pi)$.  Moreover, in Fig.~\ref{fig:kxkyandslices}(e) the real part of the Green's function for the antinodal cut shows that the reduction of states coincides with a dip in ${\rm Re}G^{(4)}$ to take on negative values at $(0, \pi \pm \delta)$. 

There is a long history of phenomenology of pseudogap physics and its connection to Fermi surface reconstruction that extends from the the mean-field level\cite{yrz,yrzreview,leblanc:yrz:raman, leblanc:njp} to exchange coupling\cite{kochetov:2018}, including also DQMC and cluster-DMFT methods\cite{trivedi:2020,ferrero:2018:fs}.  In the latter case, all previous calculations have been restricted to analysis based on numerical analytic continuation schemes of Matsubara axis data or indirect observation from other quantities.  
In our results, we see directly for the evaluation of the spectral density in real frequencies evidence of Fermi surface reconstruction.\cite{vishik:2010}  The sign change in ${\rm Re}G$  is an indicator of a change in character of states from hole-like to electron-like, similar to electron/hole pockets observed in phenomenological models.\cite{yrz,ashby:2013}
Qualitatively, our results suggest that Fermi surface reconstruction can occur at arbitrarily weak interactions and coincides with the antinodal non-Fermi-liquid crossover identified in Ref.~\onlinecite{fedor:2020} shown to originate from strong spin-fluctuations.\cite{Schaefer:2020} 
These calculations at fourth order are approximate at this value of $\beta t=5$, and while higher order results are extremely expensive, they are in principle possible within the same AMI framework.  Extending this work to higher order and improving convergence of the integration of internal parameters will be necessary to make quantitative conclusions about the nature of Fermi surface reconstruction in the 2D Hubbard model. 

\section{Dissecting the Dyson expansion}\label{sec:dissect}
\begin{table}[]
    \centering
    \begin{tabular}{c|c|c|c|c}
       Order  & n$_{\Sigma}$ & n$_{G}$ & n$_{\rm red}$ & $\frac{{\rm n}_{\rm red}}{{\rm n}_G}$(\%)  \\ \hline
       2  &1 & 1& 0 & -- \\
        3 & 2 & 2& 0 & -- \\
       4  & 12 & 13 & 1& 7.7\\
       5  & 70 & 74& 4 & 5.4\\
        6 & 515 & 544& 29 &  5.3\\
        7 & 4264 & 4458  & 194 & 4.4\\
    \end{tabular}
    \caption{Order by order counting of diagrams in the self-energy expansion, n$_{\Sigma}$, Greens function expansion, n$_{G}$, and the number reducible diagrams that account for the difference, n$_{\rm red}$.}
    \label{tab:diagram_counting}
\end{table}

In this section, we illustrate the reason for fundamentally different physics in the direct expansion of $G^{(n)}$ as compared to obtaining $G_{\Sigma^{(n)}}$ from Dyson's equation. To streamline discussion, we  consider the direct contribution of each diagram to a particular observable, the imaginary Green's function at the antinodal momentum for the zeroth Matsubara frequency.
We will see that  a pattern emerges which clearly separates the reducible contributions included in the Dyson series from the irreducible contributions that are neglected at each order.

First, we consider for the Hubbard interaction the number of diagrams in each 
expansion, enumerated in Table~\ref{tab:diagram_counting}.  After absorbing tadpole diagrams, the number of terms in the self-energy expansion is given by 
n$_{\Sigma}$.  Similarly, the number of diagrams in the Green's function 
expansion is given by n$_{G}$.  The difference between the two values is the number of 1-particle reducible diagrams n$_{\rm red}$ of that order, each of which is comprised of combinations of self-energy insertions of lower order.  It is therefore the set of n$_{\rm red}$ diagrams that are implicitly resummed at a particular order when one calculates the self-energy and then employs the Dyson equation.  For example, 
if one truncates the self-energy expansion at fourth order and then utilizes Dyson's equation, the effective contribution at sixth order is the sum of n$_{\rm red}=29$ diagrams out of a possible n$_{G}=544$ diagrams, 
only $\sim 5\%$ of the diagrams.  This disparity between the number of diagrams included in the Dyson expansion and the full number of diagrams at each order does not obviously change with truncation order.  This is illustrated in the final column of Table~\ref{tab:diagram_counting}, the value of n$_{\rm red}$/n$_G$ remains small at $\sim5$\%. 
This means that for any self energy expansion truncated at order $m$, the application of the Dyson's equation only accounts for $\sim 5\%$ of the diagrams at order $m+1$ ($m+2$ in the case of half-filling where odd orders vanish).
The fact that the fraction of included diagrams is small is not necessarily problematic so long as those n$_{\rm red}$ diagrams are the dominant contribution.  

\begin{figure}
    \centering
    \includegraphics[width=\linewidth]{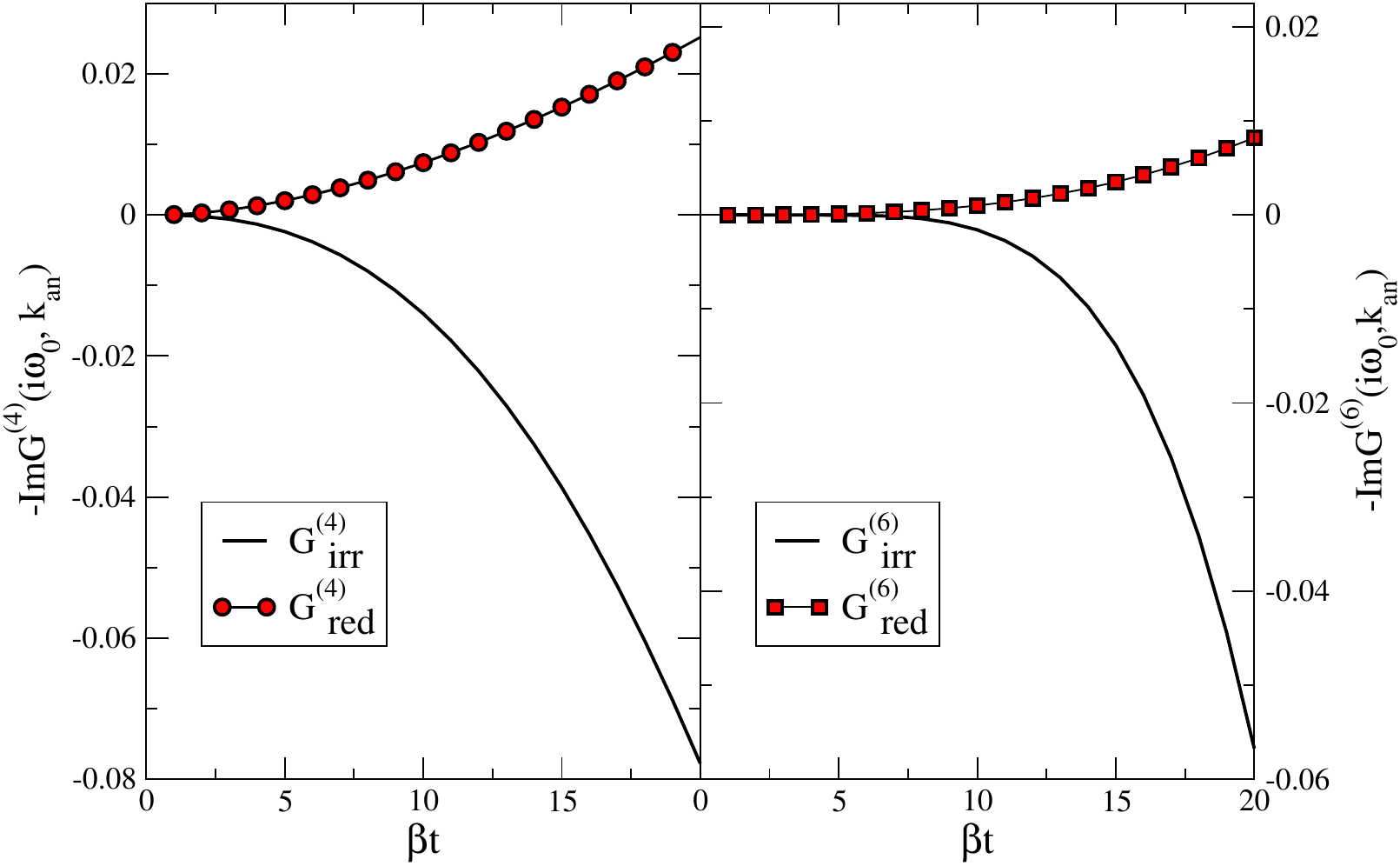}
    \caption{Evolution with $\beta$ of contributions to $-{\rm Im}G(i\omega_0,k_{an})$ from reducible and irreducible diagrams for 4th order (left frame) and 6th order (right frame).  Shown are the direct coefficients equivalent to the $U/t=1$ case. }
    \label{fig:red_irr}
\end{figure}

We probe this issue in detail in Fig.~\ref{fig:red_irr} where we plot the negative imaginary part of $G^{(4)}$ (left frame) and $G^{(6)}$ (right frame) broken into contributions from reducible and irreducible diagrams.  Shown as a function of inverse temperature for the zeroth Matsubara frequency, a positive value is indicative of metallic behavior while a negative value suggests insulating behavior. At 4th order the reducible and irreducible contributions are comparable in magnitude but opposite sign for all values of $\beta t$.  Below $\beta t=1$ the two contributions merge, but tend to zero suggesting that neither contribution is important for temperatures above $T/t=1$.  
At 6th order we again see that the reducible and irreducible contributions have opposite signs and below $\beta t \approx 6$ both contributions are negligible.  
 Since we are comparing coefficients of the same order - carrying the same factors of $U^m$ - this disparity in value is not interaction strength dependent, and exists for all choices of $U/t$. 
%We probe this in Table~\ref{tab:irrred_values} where we quote values of reducible and irreducible groups at each order for the case presented in Fig.~\ref{fig:FedorSigma} [$U/t=3$, $\beta t=8.33$, $\mu=0$] at the zeroth Matsubara frequency, $i\omega_0$. We see at each order that the reducible contributions, though accounting for a small number of diagrams have an amplitude comparable to the irreducible diagrams. However, the reducible and irreducible diagram contributions have opposite signs which in this case represents the wrong physical outcome - insulating vs metallic behavior.  
%Since we are comparing coefficients of the same order - carrying the same factors of $U^m$ - this disparity in value is not interaction strength dependent, and exists for all choices of $U/t$.   

In order for Dyson's equation to have utility, the sum of reducible diagrams would need to be the dominant contribution to the total result.   If this were the case, then the infinite sequence of reducible terms generated by Dyson's equation would provide for free a good approximation of very high order terms.  Interpreting the results of Fig.~\ref{fig:red_irr} we see that there is no temperature at which the reducible diagrams are the dominant contribution.  We see that systematically, order by order, the reducible contributions are not a good representation of the total.  This does not appear to change as order increases and we expect that, for fixed temperature, this disparity in value will continue until both the reducible and irreducible contributions at the next order are negligible.

The basic logic of the Dyson series is that it should be preferable to include \emph{some} higher order terms rather than none.  We see instead that it is systematically preferrable to \emph{not} include those terms.  
This suggests that there is no advantage to employing Dyson's equation for this problem - the Hubbard interaction at weak coupling and low temperatures.
This does not mean we should not take advantage of the reducibility of self-energy diagrams, simply that we should truncate the expansion and include all terms of each order to avoid this issue.

\section{Conclusions}\label{sec:conclude}

In this work we have presented results for single-particle quantities from partial diagrammatic expansions of the 2D Hubbard model, for both Matsubara and real frequencies, made possible by recent advancements in symbolic tools.\cite{AMI,GIT,AMI:spin} Thanks to exact results, at extremely high order,\cite{fedor:2020} we compute and contrast the direct expansion of the Green's function against the use of the Dyson series expansion. 
We compute the single particle properties of the half-filled model including up to sixth order contributions, a feat that requires the generation and evaluation of $\sim 10^6$ analytic expressions.  We evaluate the expansion for weak coupling, $U/t<4$, where we observe many expected features such as the formation of a gap in the density of states, as well as a separate behavior of nodal and antinodal Fermi surfaces which are qualitatively similar to Fermi surface reconstruction.
Motivated by the phase diagram of Ref.~\onlinecite{fedor:2020}, we set out to establish the consequences of observed non-Fermi liquid behavior on the real frequency spectral function.  In this process, we determined that for both real and Matsubara axis data, the low order expansion of the Green's function demonstrates completely different physics (insulating) from what one obtains via the self energy and Dyson's equation (metallic) for a similar order of truncation.
By dissecting the Green's function into its reducible and irreducible components, we see that the reducible diagrams that are implicitly included in the Dyson equation represent only about 5\% of the diagrams at a given order. In the case of the 2D Hubbard model, the reducible diagrams are never the dominant contribution at each order and as a result, the use of the Dyson equation creates a systematic error that is not resolved by increasing the expansion order unless all higher order terms are negligible.  We determine that this systematic error for low order expansions can be avoided by simply not employing the Dyson equation, and instead directly evaluating $G^{(n)}$ from the set of $\Sigma^{(m)}$ contributions obtained at $m\leq n$.  We show that this gives improved results with respect to our known exact benchmark. 
Since we work with only bare Green's functions, this issue, though surely related, should not be confused with convergence issues in renormalization approaches where the convergence of an iterative skeleton series converges to unphysical branches.\cite{reining:2017,kozik2015nonexistence}
Extending our work to higher order will be necessary, which will require additional algorithm development as pointed out in Ref.~\onlinecite{igor:spectral}.

In summary,  our results show explicitly that partial perturbative expansions of the self-energy based on the bare propagator when combined with Dyson's equation will lead to incorrect physical results for the half-filled Hubbard model.  
In hindsight this is perhaps not surprising since, regardless of what problem is being solved, there is no mathematical or physical reason to expect that the $\approx 5\%$ of reducible diagrams at each order that are implicitly included in the Dyson series should be a good representation of the diagrams at a given order.  This observation of the disparity in diagram numbers is however only part of the story, as explored in Section \ref{sec:dissect}.  It is reasonable to presume that  one might find a class of problems where for some limits of model parameters (eg. high temperature, low density) the reducible diagrams provide a dominant contribution which would result in correct physics from a partial self-energy expansion.  However, in absence of such a special case our observations lead us to question the general utility of the Dyson series where there are two likely scenarios for a given truncation order $n$, either:
\begin{enumerate}
\item  The $\Sigma^{(n)}$ series is converged in which case $G^{(n)}\equiv G_{\Sigma^{(n)}}$, suggesting that the additional higher order diagrams implicitly included in the Dyson series are effectively zero and the Dyson equation provides no new information.
\end{enumerate}

Or 

\begin{enumerate}[resume]
\item  the $\Sigma^{(n)}$ series is not converged in which case it is likely that the Dyson equation erroneously sums incorrect values to infinite order potentially leading to wrong physics.
\end{enumerate}

%\begin{enumerate}
%    \item The $\Sigma^{(n)}$ series is converged in which case $G^{(n)}\equiv G_{\Sigma^{(n)}}$, suggesting that the additional higher order diagrams implicitly included in the Dyson series are effectively zero and the Dyson equation provides no new information.
%   \item or the $\Sigma^{(n)}$ series is not converged in which case it is likely that the Dyson equation erroneously sums incorrect values to infinite order potentially leading to wrong physics.
%\end{enumerate}

\noindent When framed in this light, the general utility of the Dyson series is potentially compromised.  
Identifying classes of problems where the Dyson series for partial diagrammatic expansions is guaranteed to lead to correct physics could represent a fruitful direction of future study.

\begin{acknowledgments}
 JPFL and GTA acknowledge the support of the Natural Sciences and Engineering Research Council of Canada (NSERC) (RGPIN-2017-04253 and RGPIN-2015-04306). 
Computational resources were provided by ACENET and Compute Canada. Our Monte Carlo codes make use of the open source ALPSCore framework \cite{Gaenko17,alpscore_v2}.
\end{acknowledgments}

\appendix

\bibliographystyle{apsrev4-1}
\bibliography{refs.bib}

\end{document}